\documentclass[jocse]{jocseart}

\usepackage{booktabs} 

\setcopyright{jocsecopyright}

\jocseDOI{10.22369/issn.2153-4136/x/x/x }

\begin{document}
\title{Bridging the Educational Gap between Emerging and Established Scientific Computing Disciplines}
\subtitle{How to teach scientific computing to students in medical sciences and other emerging fields}

\author{Marcelo Ponce}
\orcid{0000-0001-5850-7240}
\email{mponce@scinet.utoronto.ca}
\author{Erik Spence}
\email{ejspence@scinet.utoronto.ca}
\author{Ramses van Zon}
\email{rzon@scinet.utoronto.ca}
\author{Daniel Gruner}
\email{dgruner@scinet.utoronto.ca}
\affiliation{%
  \institution{SciNet HPC Consortium, University of Toronto}
  \streetaddress{661 University Ave.}
  \city{Toronto}
  \state{ON}
  \postcode{M5G 1M1}
  \country{Canada}}
\affiliation{%
  \institution{Institute of Medical Science, Faculty of Medicine, University of Toronto}
}

\renewcommand\shortauthors{Ponce, M. et al}

\begin{abstract}
In this paper we describe our experience in developing curriculum courses aimed
at graduate students in emerging computational fields, including biology and medical science.
We focus primarily on
computational data analysis and statistical analysis,
while at the same time teaching
students best practices in coding and software development.
Our approach
combines a theoretical background and practical applications of concepts.
The outcomes and feedback we have obtained so far have revealed several issues:
students in these particular areas lack
instruction like this
although they would tremendously benefit from it; we have detected several
weaknesses in the formation of students, in particular in the statistical foundations
but also in analytical thinking skills.
We present here the tools, techniques and methodology we employ while teaching and developing this type of courses.
We also show several outcomes from this initiative, including potential pathways for fruitful multi-disciplinary collaborations.
\end{abstract}

%
%

\ccsdesc[500]{Social and professional topics~Computing education}
\ccsdesc[500]{Social and professional topics~Model curricula}
\ccsdesc[500]{Social and professional topics~Student assessment}
\ccsdesc[300]{Social and professional topics~Computational thinking}
\ccsdesc[300]{Social and professional topics~Computing education programs}
\ccsdesc[100]{Social and professional topics~Accreditation}
%
%

\keywords{Training and Education, Computational Statistics, graduate courses, curricula, student assessment}

\maketitle

\section{Introduction}
\begin{figure}
	\includegraphics[width=\columnwidth]{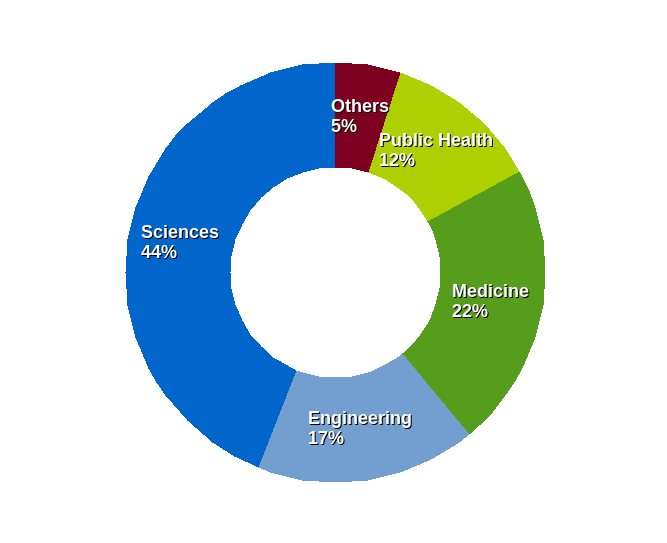}
	\caption{Distribution of student's departments/institutions in all SciNet courses, period 2013-2017.
		As can be seen traditional HPC disciplines, such as STEM, still constitutes the majority (roughly 60\% of our trainees).
		However, biology and medical students are quickly catching up in numbers representing approximately 35\% and with an increasing trend in the last years \cite{website:SciNet-Training}.}
	\label{fig:distribStudents}
\end{figure}

In this paper, we present the methods and strategies we have used to
expand our traditional scientific and high-performance computing
programs into university-curriculum courses for disciplines ranging
from physics and biology to medical sciences
\cite{DBLP:journals/corr/PonceSGZ16,Roberts:2011:MCR:2003616.2003617}.
We show our steady growth in these disciplines, demonstrating a clear
need for approaches like ours, not only for the traditional
high-performance computing sciences but also for the not-so-usually
engaged disciplines, as shown in Figures~\ref{fig:distribStudents} and
\ref{fig:2012-2017_attendance_hours}.  We discuss the methodology we
use to teach these non-traditional students. 

This paper is organized as follows: in Section~\ref{sec:motivation} we
explain the main motivations and goals that we target when designing a
course like this; Section~\ref{sec:courseDesign} describes the basic
layout and main elements of the course;
Section~\ref{sec:methods} describes the methodology we use to evaluate
and transfer the knowledge to the students, including the different
strategies have we used for offering the course; in
Section~\ref{sec:results} we evaluate the outcomes and discuss future
directions and implementations for the course; in
Section~\ref{sec:conclusions} we draw some final reflections about the
approach presented in this paper.

\section{Motivation}
\label{sec:motivation}
Proper training in data analysis and statistical
techniques is indispensable for modern scientific research.  Recent
times in particular have seen the adoption of computerized data
analysis techniques in all fields (biology, human sciences, medicine,
\emph{etc.}).  The last decade has also seen the advent of a new era
of data availability and scale, with huge amounts of data easily
collected and shared among scientific researchers, governments and
businesses.

However, the skills and knowledge needed to seize the opportunities
presented by these data have, in general, not been taught in university
courses.  Researchers, in particular graduate students and postdocs,
are largely forced to learn these skills on their own, if they learn
them at all.  The effort to understand basic concepts and overcome the
technical difficulties associated with data analysis tools diverts
from and delays the main goal, research.

Expecting students to pick up this knowledge and skill by themselves is
especially troublesome in fields that do not have a tradition of doing
computational research, such as biology, medical science, health
science, humanities, \emph{etc.}  Students in those fields cannot
turn to their senior colleagues for guidance, something that is common
in traditionally computational disciplines like physics, chemistry,
engineering and astronomy.  We have offered scientific computing
courses for students in the physical sciences for many years, but
these courses focused on compiled languages, numerical libraries,
solving differential equations, and parallel computing. They were
found to be ill-suited for most students in biomedical fields both
because of required prior knowledge and the mismatch of topics
covered. Biomedical computation is more likely to involve statistics,
data analysis and interpreted languages, not discretized partial
differential equations.

Our recently developed courses aim to fill this gap
by training students in the practical
application of statistical data analysis, machine learning tools, and
professional coding practices.  It begins by offering an
introduction to the R programming language \cite{Rlanguage},
an open source tool for data analysis
that is popular in the medical sciences. 
Basic concepts and elements of statistical analysis
are presented, not only by reviewing theoretical foundations but
also by examining examples and applications.  Next, statistical methods such
as hypothesis testing, parametric and non-parametric model creation,
model diagnosis, clustering and decision trees algorithms are
discussed and implemented using R, and applied to several
real-world examples.
This particular course is aimed to graduate students (master and doctorate degrees)
from the Institute of Medical Science (IMS) at the University of Toronto.

In previous years, we offered a very successful subset of the current course,
in a modular format. One important conclusion we drew from students' feedback and comments
was that the students would definitely benefit from a more comprehensive and extensive
program, incorporating more advanced topics and extending the duration of the course.
Hence the latest iteration of the course spans a full semester.
The course is in high demand:
in the last year (fall 2017, winter 2018) we delivered
this course in two consecutive terms and the registration was above the original number of spots
reserved for the course, resulting in the creation of a long waiting list each time
and an increase of 33\% in the planned size of the class for the next year.

\section{Course Design}
\label{sec:courseDesign}
The goal of this class is to prepare graduate
students to perform scientific data analysis.  Successful students
 learn how to use statistical inference tools to gain insight into
their data, and are exposed to cutting-edge techniques and best
practices to store, manage and analyze data.
We use the R Statistical Language \cite{Rlanguage}
to teach the students the basics of programming and how to perform
proper data and statistical analysis.

We focus on four main areas in the course: i) computer programming techniques, including:
 basics of programming, functions and arguments, documentation and well commented codes and scripts;
 ii)
software development 
best practices, such as, modularity, version control and proper file IO operations;
iii) implementation of statistical analysis techniques and pipelines employing the programming skills
transferred in the previous points, \emph{i.e.}~computational statistics;
and iv) advanced and cutting-edge statistical analysis, including machine learning and neural network implementations.

By the end of the course, the students should have developed basic
programming skills and created a set of tools and scripts that
help them analyze and tackle their own datasets and research problems.

\subsection{Course Content}
The course is given in twelve weeks, with two 1-hour lectures per week. Grading is based on 10 assignments set throughout the course, and further discussed below.  The course is open to all graduate students at the university but there are a limited number of spots (currently 60), that were all filled in the past.

The list of topics covered includes:
\begin{itemize}
\item Introduction to the Linux Shell.
	File manipulation, regular expressions, bash scripting,
	automation of data-analysis pipelines.
\item Introduction to programming with R.  IDEs and R
  standard console, basic programming concepts: conditionals, loops,
  variable types.
\item Introduction to programming best practices in R.
  Functions and scripts, interactive versus batch processing, variable
  scope, modular programming, defensive programming, comments and
  documentation.
\item Introduction to version control.  Motivation,
  implementation and use of version control, GIT, logs, rollback, branches.
\item Binary file input/output.  Accessing, reading and
  writing binary data, file input and output strategies, and best
  practices.
\item Basic statistics using R.
	Review of the basic concepts of probability and statistics, probability distributions,
  descriptive and inference statistics.  Statistical modelling
  implementations using R: linear models, quadratic models,
  generalized linear model.  Testing models: correlation and
  covariance, Pearson coefficient.  Hypothesis testing: examples of
  null hypothesis tests implementations, T-Student test, two sample
  T-test, matched pair experiments, independence tests, ANOVA-like
  based tests.  Model Diagnostics: graphical tools, leverage,
  influential points, Cook's distance, residuals, validation of
  assumptions.
\item Advanced statistical topics: Generalized linear models,
  power analysis, survival analysis, structural modelling equation, etc.
\item Statistical discussion of some important ``paradoxical'' cases:
	Simpson's paradox, Anscombe's quartet, ...
\item Introduction to machine learning.
	Regression, overfitting, bias-variance tradeoff, cross-validation,
	bootstrapping, LOESS, LOWESS.
\item Advanced machine learning.
	Variable selection, dimensionality reduction, principal component analysis.
\item Classification algorithms.
	Decision trees, confusion matrices, clustering, logistic regression, Naive Bayes.
\item Introduction to Neural Networks.
	Motivation. Basic examples and implementations in R.
\item Visualization of data.  Publication-quality figures,
  basic plotting, 1D (curves), 2D (contour maps, heatmaps, dendograms, etc) and 3D plots,
  interactive visualization, animations.
\item High-performance R.  Memory management, in-core
  processing, byte-compiling, C++ interfaces, parallel techniques.
\end{itemize}

Examples and assignments, presented and discussed within the course, cover
study cases based on clinical trials, drug tests, medical cases and hospital treatments,
differential gene expression, bioinformatics and *omics techniques, etc.

\subsection{Prerequisites}
Students should ideally have some light programming experience in any language,
and a bit of command-line experience is a plus. 
Students should have a laptop to bring to the lectures, with R installed,
which is freely available for Linux, OS X and Microsoft Windows.
We have noticed that due to the way we are able to deliver the course even
students with no previous experience in coding are still able to follow
and succeed in the course; however their dedication and time commitment 
might be a bit higher than for other students with a background in computing.
Initially, we assumed that because this was aimed as a graduate course,
students would have taken previous courses in statistics, however not all of them
have a solid foundation or have even taken a recent course on basic statistics.
Therefore, we decided to add as prerequisites for the course some basic knowledge on
and exposure to at least one statistical course.

\subsection{Passing Requirements \& Grading Scheme}
Most weeks, students are given a programming assignment, with a due date one week after.
These assignments are designed to help students absorb the course material.
There are 10 assignments in total.
The average of the assignments makes up the final grade.
To ensure a timely reporting of student grades, we adhere to the following policy:
homework may be submitted up to one week after the due date, at a penalty of 0.5 point per day,
out of the 10 points for each homework assignment.
All sets of homework need to be handed in for a passing grade, although
a make-up assignment can be given at the end of the course. Rather than focusing on the topic
of a specific week, the make-up assignment may involve any of the material covered in the course.

Attendance is not mandatory for the course, but strongly recommended.
This constitutes an important departure with respect to other courses
offered at the university level, in particular for IMS students.
Because the way we deliver our lectures and we offer the material to students (see next section)
students have the flexibility of attending or watching our classes remotely.

\section{Methodology}
\label{sec:methods}

\subsection{Strategies}
\label{sec:strategy}

One of the main challenges of formally implementing and offering this
course was to make it available to students across the university as
a listed graduate course. The difficulty originates from the fact that
SciNet, the supercomputer department of the University of Toronto and
the home department of the instructors, is not a teaching department.

One of the most efficient ways we found to overcome this was
partnering with other departments or institutes at the university level
(\emph{e.g.}~the Institute of Medical Science, Physics Department, Department of Physical and Environmental Sciences).
By doing so, we provide a formal framework for the course, allowing students to enroll
through the official university system, thus being recognized in their official transcripts,
\emph{i.e.}~taking the course for actual university credit.
At the same time, by having the course listed on the official calendar, the course
is also visible to students from any other departments at the university,
thus increasing its exposure to the graduate students.

With respect to strategies related to the consolidation of knowledge and application of concepts,
an approach we usually employ in some of the assignments is to ask the students to \textit{use their own data},
if it is the case that they have data which is suitable and available to be used in the assignment.
This approach has several advantages, on the one hand, it allows the students to make direct contact with
the techniques described in class and immediately apply them to their own research fields.
It enables students to use the tools in a more friendly and familiar environment, as it is basically
their own research questions and problems.
It also demonstrates to the student the efficiency and capabilities of the techniques and tools
we teach, and how they can be properly applied to their own research and problems.
It has the tremendous advantage of allowing us to gain some insights of what the students struggle with in their
day-to-day work, what type of questions they are trying to answer, in short what their research is about.
Moreover, we may be able to help students develop actual tools that they can then use and
bring into their corresponding labs and groups.
The only downside to this sort of assignment, as these are quite open, is the grading itself.
In this case we don't have tentative solutions that we can offer to the TAs (see below), 
but we provide very detailed guidelines.
Similarly, we also ask students to focus on particular techniques we discussed in class.  Although they are welcome
to use others not introduced in the course, they need to explain those in a brief report following a traditional paper structure,
which also constitutes part of this type of assignment.

In a similar vein, we ask students to create scripts
that allow them to produce professional/publication quality figures.
At the same time that we evaluate students' understanding of the material,
we empower them by creating tools that will be of utility when they
need to generate plots for their own papers or research projects.
To show off some of the remarkable results we have obtained, we have created a
\textit{Visualization Gallery} \cite{website:viz-gallery} where we post and display
figures, graphics and animations created by the students of our courses.

\label{sec:TAs}
Another crucial point that touches upon the partnership with other departments
and institutes is the necessity of having \textit{teaching assistants} (TAs)
helping with the grading of the assignments.
The importance of having TAs has been shown in many situations, \emph{e.g.}~\cite{roberts1995using}.
In particular for courses with this number of students (approximately 60 per term)
and 10 assignments per course, meaning around 600 assignments to grade.
Furthermore, the way we grade the assignments is not automatically or in bulk, \emph{i.e.}~we do look at each assignment individually, look at the logic of the implementation,
whether it works, if it's logically correct, and we provide individual feedback and 
detailed comments for each student.
This is quite labour intensive, and hence not doable without the support of the TAs, which
are the main graders of the course.
The fact that the TAs are financially covered by the departments sponsoring the course,
not only allows us to provide speedy turnover in the comments and feedback to the students,
but also to substantially increase the number of students we can accept into our courses.
For a course with 60 students we utilize 3 TAs, who are current MSc or PhD students.
Ideally we expect the TAs to have experience with programming in general and specifically in R,
Linux OS and command line terminal, version control systems (such as GIT), and
being knowledgeable in statistics.
Not surprisingly, excellent candidates for these TA positions are often students who have taken our course
before, as they have being trained on the good practices we want to emphasize,
have received the type of feedback we want them to provide to fellow students, and
they have experienced first hand the course and the whole evaluation process.

The TAs' main duties are marking and grading 10 weekly assignments,
i.e. evaluate assignments and provide comments and feedback.
We also recommend that the TAs attend and review the content presented
in the lectures (12 lectures at 2 hours per week) and spend some time in preparation for the grading
(review topics covered in class and get familiar with the material).
We usually host weekly meetings with instructors to discuss assignments and grades.
In general, depending on the department a workload like this is given between 70 and 120 hours of work per TA.

\begin{figure}
	\includegraphics[width=\columnwidth]{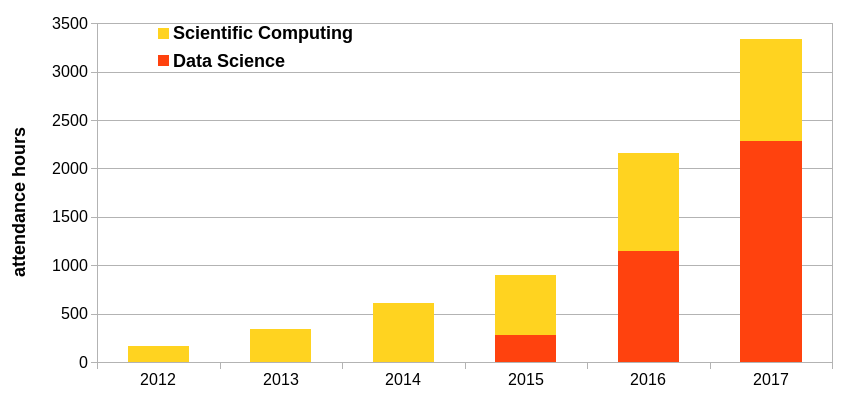}
	\caption{Aggregated attendance hours on ``Data Science'' and ``Scientific Computing'' graduate courses in the period 2012 to 2017.
		The effects of size increment, mostly due to the ability of being able to handle more students
		can be clearly seen in the years 2016 and 2017.}
	\label{fig:2012-2017_attendance_hours}
\end{figure}

As mentioned before, neither a fast feedback nor the size of the classes
would have been possible without the support of the TAs, and this can be easily seen in Figure~\ref{fig:2012-2017_attendance_hours};
each time a new partnership has emerged there is a bump in our enrollment numbers.
However to make things work smoothly there is a lot of logistical work that has to be done,
ranging from the usual paperwork for hiring and selecting candidates, up to 
the most important part related to the grading itself.
In particular, we do provide the TAs explicit guidelines, grading schemes and tentative solutions
for each assignment. We try to anticipate and cover all the basic mistakes and important
concepts the students could face on the assignments. This is a lot of work upfront, but
it is totally worth it for the number of students we have to handle.
Of course, sometimes we have to look at a particular student's submission
or assignments, as there are always corner cases, but this is still a manageable amount.

\subsection{Lectures}
The type of knowledge we teach in our classes is rarely
found at university-level courses and is quite sought-after.
In many cases and disciplines, students and even researchers spend a good
amount of time self-learning many concepts we include in our ``best practices'' topics,
such a modularity, testing and developing strategies, defensive programming, version control, etc.
To instill such best practices in a graduate course setting,
we have found that one of the most efficient ways to deliver
our lectures is with a combination of theoretical concepts and practical examples and applications.
Depending on the particular topic to be presented, we emphasize a more
practical/hands-on approach during the class -- for instance, when covering topics such as
introduction to Linux Shell, exploring the basics of the R language, visualization, \emph{etc.}.
However, if the topic to be covered requires some theoretical background,
we then deliver a more traditional lecture --
\emph{e.g.}~reviews on probability and statistics, introduction to machine learning and neural networks, etc.
In general, even when we are discussing and presenting theoretical concepts
and examples, we still mix practical implementations of such concepts using the computer,
so that the students can actively follow the computational implementation during the lectures.

We usually face the question whether we would prefer to deliver our lectures
in a computer-lab or traditional classroom.
Having a computer-lab based classroom offers the advantage of controlling
the workstations and computers, as well as the installation of the software and libraries
to use.
However, we prefer that the students use their own laptops instead, so that
they actually have to go through the process of installing, fixing and sometimes troubleshooting
problems dealing with their own computers and installations.
This comes back to the point that our courses are aimed to deliver practical skills,
so that the students sitting in our classes can help others in their labs and groups, 
after having transited that path before with our guidance and support.
There are some features that are desirable in traditional lecture rooms:
spacious locations are preferable, so that students can comfortably place their laptops
and there is room for the instructors to help and move around, especially
for the lectures when there is an important hands-on component.
As students will be following along, features such as power bars and outlets are desirable, as well as
a reliable wireless internet connection.
Our lectures are delivered mostly using slides so having a projector is crucial, but
often involve the blackboard as well
to clarify or demonstrate
some concepts or examples.
For big rooms, having the support of AV systems, in particular microphones and
speakers is important as well.

\subsection{Tools}
\textbf{Online resources}:
All our material (slides, assignments, lecture recordings) is freely available online in our \textit{education website} \cite{website:SciNet-Edu}.
We keep all our past courses, lectures, slides, assignments and recordings available on their respective websites\footnote{See, for instance, hyperlinks for our previous editions:
\href{https://support.scinet.utoronto.ca/education/go.php/262/index.php/ib/1//p_course/262}{Fall 2016, first module version} part of the Translational Research program,
\href{https://support.scinet.utoronto.ca/education/go.php/324/index.php/ib/1//p_course/324}{Fall 2017, first full course edition},
\href{https://support.scinet.utoronto.ca/education/go.php/359/index.php/ib/1//p_course/359}{Winter 2017/2018, second full course edition}.
}.

Our educational website \cite{website:SciNet-Edu} is developed based on
the open source web-based learning content management system
ATUTOR \cite{atutor}.

One resource that students appreciate very much is the fact that we \textit{record}
all our lectures and post them on the course website.
This allows students to access the material even if they were not present
in the class or if they want to review and revisit some of the topics covered in class.
In some cases we also \textit{live stream} the lectures if there is demand,
with a live chat to answer remote questions
as supported by the ATUTOR\cite{atutor} framework.
For recording and streaming the lectures we use a set of mostly
open source tools\footnote{Open Broadcaster Software: \url{https://obsproject.com}, Camtasia: \url{https://www.techsmith.com/camtasia.html}, ActivePresenter: \url{https://atomisystems.com/activepresenter}.} 
 in combination
with our own setup for streaming video \cite{videowall-in-prep}.

Another useful and practical online resource, that we offer is an
online \textit{forum} system (also provided by ATUTOR), where the students can post questions,
see questions posted by other students and even answer the questions, in other words,
start a conversation among their own peers. We, the instructors and TAs, also keep
an eye on it, and try to answer their questions there too.
One nice feature that the system offers, is that the users can register to get email
notifications when new posts are added in the forum.

Another important online contact resource we offer to the students is \textit{email},
this is by far the online resource most used by the students.
For instance, in this last year's edition alone, we have answered around 4000 emails.
It is quite challenging to keep this under control and balance, but we
know this way of interaction is greatly appreciated by the students.
We personally, as instructors, like to see and answer every email,
as that also provides us with important insights and diagnostic information
about what topics the students are struggling with the most.
In some cases, when we detect some particularly problematic issues, we can take quick
measures to help alleviate the problems. For instance, we can clarify questions
or address particular points during the classes, and even in some cases if it is
more of a technical issue, we would create a post or discussion item in the
course website.

We have also found that a crucial part of the learning process and something students
truly appreciate are our weekly \textit{office hours}, which students can attend
to pose questions either on particular topics covered in class or to 
get help while they work on their assignments.
Not surprisingly, we have found that just one hour is not enough and
we usually find ourselves extending the period or even staying for around 30 minutes
after class discussing and answering questions from students.

\subsection{Evaluation Methods}
Our main evaluation avenue in the course are quasi-weekly assignments.
The reason we prefer assignments over mid-terms and/or finals is
because we think that having almost one assignment per topic covered
offers us the possibility of evaluating with much more fine granularity
the knowledge gained by the students.  It also offers the students
the opportunity to practice the concepts discussed in class.
Moreover, the type of knowledge we try to transmit in our lectures is applied/practical
by its own nature. Thus having the students implement something by themselves
is the best scenario we could think of to reinforce the learning and concepts
presented in class.
The assignments are designed with two major objectives:
 1) to offer the student a chance to practice the most
relevant parts of the techniques or concepts discussed in class, and
 2) to challenge the students to digest and think
beyond the material that was presented, presenting problems in which they need to join different
techniques to arrive at new results.
Usually we like to set up the assignments with a sort of ``hidden message'',
a learning opportunity, something the students can discover by themselves by following
specific guidelines and clues that we leave for them. In this case we believe this self-discovery
process is much richer than the knowledge one can transfer in any sort of direct or explicit message
delivered in lectures.

This way of evaluation posses quite significant challenges \cite{parlante2006nifty}: coming up with actual assignments
that fulfill such a role, find the suitable sweet spot of being interesting but not too hard
to overwhelm the students, and still be amenable to grading.
However one of the most difficult challenges to take into consideration when
having this type of homework is being vigilant of students sharing solutions
or working on the same code/submission, in other words any sort of plagiarism.
We encourage students to openly discuss with peers,
but we strongly enforce individual work. No collaborative work or submission is allowed
under any circumstances. We strictly follow  the university's
``Code of Student Conduct''\footnote{\url{http://www.governingcouncil.utoronto.ca/Assets/Governing+Council+Digital+Assets/Policies/PDF/ppjun011995.pdf}}
regarding plagiarism.
In this particular case, this challenge is even harder \cite{roberts1998strategies,roberts2002strategies},
as the students are submitting programs, scripts, in many cases just pure code.
So in order to tackle this issue, we have in place a series of tools.
When possible we actually run some scripts we developed with the goal
of identifying substantial overlap in the submitted assignments (similar tools exist for students when
writing essays or papers --see \cite{turnitin}--).
If there is just one TA grading a whole batch, usually in smaller class sizes (upto 20-30 students),
then we ask the TA to be vigilant about this type of situations and can warn us about any suspicious submissions.
In the case of larger courses, we can basically take two approaches:
i) Have one TA doing the grading for one whole assignment; this has the benefit that
this one TA can see all the submissions at once and identify any potential overlapping assignments.
It also helps to normalize the grading criteria -- even when we provide concrete and precise
guidelines and grading schemes, each TA has also his own style,
a fact reflected in the assignment feedback.
However, this approach has its own disadvantages too, as it takes longer to
give feedback to the students, and it is also more sensitive to grader bias.
ii) A second approach, which is the one we follow for larger courses, is to equally
divide the number of submissions among the TAs and have all of them working in parallel
grading a subset of the assignments. This method allows us to achieve impressive turnaround times,
of no more than 48 hours! \emph{i.e.} on average two days after the students had submitted their assignments
they got feedback, comments on what they did correctly and what to improve for future assignments.
In order to minimize biases here, we randomize the list of students to be graded by the TAs.
The disadvantages of this approach are: that the grading score is not quite normalized
as there are different TAs grading at the same time (we believe that this is
a really minor point, mostly because the precise grading instructions we provide to the TAs leaves
little room for that, and if any particular issues are noted we are ready to intervene);
and secondly, it is harder to catch situations as the ones mentioned before, however we have still been
able to identify and detect cases of suspect plagiarism. 
After dealing with the incidents in question,
the subsequent assignment submissions from those students are graded by the same TA to prevent recurrence.

In other courses we also use some online quizzes that allow us to quickly evaluate,
with multiple choice questions, basic concepts the students should assimilate.
The procedure is completely automated and included in the features of the
ATUTOR \cite{atutor} web-platform that we use in our education website, hence
giving us rapid access to results and diagnostics.
This particular technique, due to its automated nature, can be easily implemented
in courses with large numbers of students.
A variation of this technique had been employed in the
shorter precursor of the current course, where
it was used to take attendance live during class with minimal disruption,
with approximately 100 students in class.

Finally in more-advanced courses, we also employ research based projects,
which include having the students working on a particular project, submitting
a preliminary report, a final report and a presentation describing the project
to the rest of the class by the end of the course.
This technique even while quite powerful and interesting, is more desirable and
applicable to more mature students, having solid foundations and clear understanding
of what the goals of the project should be.
As the projects can grow in complexity and significantly change as the projects evolve, these are sometimes
quite close to actual research explorations the students are pursuing (e.g. \cite{Berger:2018aey}), hence one needs to
closely follow the evolution of the students and the projects.
Because of this very same reason, this technique is probably not suitable for large classes,
and if the class size is above the desirable number of projects/students to follow,
or if the project appears to be too complex, partnering students in groups could be a good
way to accommodate those situations.

\section{Discussion}
\label{sec:results}

\subsection{Outcomes}
There are several ways in which we can aggregate the outcomes from this type of courses.
From observations during the course, assignment evaluations, and 
 interacting with students
during the office hours, we were able to detect a few weaknesses related to some particular areas.
For instance, we noticed that in some cases, beyond the obvious differences in academic formations and backgrounds,
there are some serious weaknesses that academic programs could and should target in incoming students.
Among the most concerning are weak analytical and critical-thinking skills,
and insufficient mathematical and statistical foundations and ability to understand concepts.
This is of course worrisome but clearly provides some important information and
indicators for academic program designers to take into consideration.
From a more technical side, one of the most challenging topics
for the students to assimilate is the concept of \textit{functions},
arguments and return statements.
Students were able to understand modularity, and even functions as elementary blocks in
a modular framework, but the passing and receiving of arguments and/or returning
information from functions into new parts of the code was probably the hardest
concept many of the students dealt with.
Of course, it is arguable that this could be one of the expected struggling points
due to the abstract nature of the concept.
However, this is also very important information for our future editions of the class.

Another interesting observation is the effect of ``early dropouts''. These are students that dropped out of the course very soon after it started, between the first
and second week, sometimes even before the first assignment was posted.
These early dropouts constitute not more than 10\% of the class size, which in practice
does not pose a problem (the enrollment was so high that we 
had a waiting list for students that did not initially get a spot).
Interestingly, we speculate that this effect can be mostly due to a couple of causes:
i) either the course was not what the student was expecting;
ii) the workload demanded by the course might have produced a negative impression on the student (however as we argue later in this section it is quite the opposite);
iii) the course structure was not appealing to the students.

One quantitative indicator of how well the course is perceived among students
is course evaluations run at the university level.
These surveys are optative, and the students
can decide to complete them providing anonymous feedback about the course.
The questionnaire has two major components: a series of standardized questions
where students can pick numeric values ranging from 1 to 5,
representing 1:~``Poor'', 2:~``Fair'', 3:~``Good'', 4:~``Very~Good'', 5:~``Excellent''.
The second part of the evaluation is composed by open-ended questions that allow students to provide more
detailed feedback about the instructors, the level of assistance during the course,
and overall quality of the course.

The first remarkable result is the level of participation in theses surveys,
especially considering that these are optional.
The percentage of participation in our courses' evaluations is usually between
	57\% and 72\%.

Additionally for the so-called \textit{Institutional composite} questions, which include items
such as intellectually stimulating course, deeper understanding of the subject matter,
learning atmosphere and overall quality of the learning experience,
the resulting
	mean from the evaluations is
	4.5/5.0\footnote{The detailed statistical indicators are: mean=4.5, mode=5.0, standard deviation=0.22 over a period of 2 years --2017/2018--.}.

Another interesting and somehow surprising result is how the students perceive
the workload of the course compared to other courses:
	3.33/5.0\footnote{In this case the numeric scale is interpreted as follows:
		1:~``Very~Light'', 2:~``Light'', 3:~``Average'', 4:~``Heavy'', 5:~``Very~Heavy''.};
which represents just a bit above the average workload.
In principle one could think that having roughly weekly assignments for almost
the whole duration of the course would be a stumbling block, however the students
realize that overall, comparing the stress of just one or two instances of evaluations
(\emph{e.g.}~mid-terms and/or finals) versus a more gradual evaluation, the latter is comparable or even preferable.

In addition to this, we could literally fill pages with testimonials about the level
of instruction, support and professionalism, these courses offer, unfortunately we
don't have a good way to measure this rather than just through anecdotal notes.

But perhaps the most outstanding quantitative result is that 97\% of the students got 
an ``A'' (i.e. their average was above 85\%).
This again provides support for an evaluation approach based on assignments.
As argued before, not only does this allow students to digest the material and implement it in
a practical fashion, but also in the end helps them learn and assimilate this knowledge,
which is a more ``fair'' and useful measure of success in the course.

One really interesting byproduct of these courses, is the potential for establishing
research collaborations among different groups and labs in need of more robust
scientific computing implementations. This is particularly true among non-traditionally
computational scientific disciplines like the ones this course targeted.
During the course,
many students approached us with open problems and potential
avenues for collaborations.
One of the most recent demonstrations of this, is an ongoing research project where
in collaboration with microbiologists and biochemists, we developed a
bioinformatics pipeline employing traditional HPC resources and open source tools,
which we expect will produce at least three publications,
the first one being already published \cite{Saettone2018}.

Another demonstration of the success of this course, is the remarkable interest
shown by the students in becoming TAs for next year's course.
By the end of every single term, we had students approaching us, asking about the
possibility of TAing for the course in future editions.
Somehow, students with a natural inclination discovered
the enchanting realm of scientific computing and seek the opportunity of becoming active participants
in the field.
As educators, this kind of outcome is just priceless.

\subsection{Future Directions}
In order to further improve our teaching and student learning experience, we
continue to develop new ideas and avenues to facilitate the transfer of knowledge
and consolidate the assimilation of basic and foundational concepts.
One way we think we can help students digest and familiarize themselves with new or difficult-to-assimilate concepts is
the development of customized lecture notes for the course, in addition to the already available slide decks. We have noticed that it is quite challenging to find resources
that we can refer the students to for further reading, as on the one hand it is difficult to find references that cover the variety of
topics we tackle in this course and with the depth we try to achieve; and on the other hand the wide range and heterogeneity in the
students' backgrounds make it even more challenging.

We are also considering implementing an additional evaluation requirement for passing the course,
consisting of an online quiz to be carried out by the middle of the semester.
The weekly assignments will continue being our main source of evaluation,
however we have noticed that students either by a lack of comprehension
or not performing the exercises in a mindful way, sometimes miss important concepts.
The in-class multiple choice quiz will help us further
diagnose difficulties in specific topics/areas and also make the students aware
of their own weaknesses.
We have implemented similar evaluation procedures for other courses, and we find
them easy to implement and evaluate using the online platform \cite{atutor} we use for our education website.

\section{Conclusions}
\label{sec:conclusions}
In this manuscript we described the road we followed in order to create
a graduate course aimed for non-traditional scientific computing students.
We believe the strategies, partnerships and methodologies we present here
can be useful for others to bridge the gap between
traditionally computational disciplines and disciplines that are new to computing.
Our approach is also different from the traditional standard university courses,
but has proven to be successful in reaching and providing new and useful tools to
students, scientists and researchers.
Furthermore, having the chance to directly interact with students we were able
to identify some important concepts that students were missing and
diagnose some crucial weaknesses which graduate programs should tackle.
Last but not least, providing this type of courses, not only offers benefits
to the students learning new skills, but it is also a way to catalyze and
push frontiers in new multidisciplinary research fields, instigating in this
way collaborations that might not have been possible otherwise.

\section*{Acknowledgements}
We want to thank all our colleagues at SciNet, and the many departments
that have partnered with us at the University of Toronto:
 Department of Physics, Department of Physical and Environmental Sciences at UTSc,
 and the Institute of Medical Science (IMS).
Special thanks to the IMS, for allowing us to offer this course, and in particular
 Prof. Howard Mount, Michelle Rosen, Sarah Topa and all the members in the
IMS' Curriculum Committee, for their confidence and continuous support
during these last 3 years of working together.
Last but not least we want to recognize and thank our amazing TAs: 
 Ricardo Harripaul, Sejal Patel, Parnian Pardis and Jonas Osmann,
for their outstanding work and commitment to the course.


\bibliographystyle{ACM-Reference-Format}
\bibliography{refs}

\end{document}